\documentclass[aps,pra,twocolumn,groupedaddress,showpacs]{revtex4}

\usepackage{graphicx}
\usepackage{dcolumn}
\usepackage{bm}

\begin{document}


\title{Bridging visible and telecom wavelengths\\with a single-mode broadband photon pair source}

\author{C. S\"{o}ller}
 \email{christoph.soeller@mpl.mpg.de}
\author{B. Brecht}
\author{P.~J. Mosley}
\author{L.~Y. Zang}
\author{A. Podlipensky}
\author{N.~Y. Joly}
\author{P.~St.J. Russell}
\author{C. Silberhorn}
\affiliation{%
Max Planck Institute for the Science of Light, G\"{u}nther-Scharowsky-Strasse 1, 91058 Erlangen, Germany
}%

\date{23 March 2010}

\begin{abstract}
We present a spectrally decorrelated photon pair source bridging the visible and telecom wavelength regions. Tailored design and fabrication of a solid-core photonic crystal fiber (PCF) lead to the emission of signal and idler photons into only a single spectral and spatial mode. Thus no narrowband filtering is necessary and the heralded generation of pure photon number states in ultrafast wave packets at telecom wavelengths becomes possible.
\end{abstract}


\maketitle

Photon pair sources are an integral part of many applications in quantum optics. Two of the most intriguing examples are quantum computation and quantum cryptography schemes \cite{O'Brien2007,Gisin2002}. While some implementations require the emission of one and only one photon pair, for example, to herald single photons \cite{Rohde2005}, others specifically take advantage of higher photon numbers, for example, to create Schr\"odinger cat states \cite{Ourjoumtsev2007}. Thus a photon pair source that can provide pure heralded single photons, while at the same time offering the opportunity to generate higher photon number states, will be beneficial in numerous ways.

In order to implement an adequate source for a certain application, it is crucial to be able to control the spatial and spectral properties of the created pair state to a high degree. Two well-established methods for the generation of photon pairs are spontaneous parametric downconversion (SPDC) in nonlinear crystals and spontaneous four-wave mixing (SFWM) in fibers. The advent of fiber sources and nonlinear crystal waveguides has simplified the control of the spatial degree of freedom by offering the possibility to restrict emission to a single spatial mode \cite{Rarity2005,Chen2006,McMillan2009}. However, in order to gain a complete description of the generated state, the spectral degree of freedom has to be considered as well. The processes of SPDC and SFWM typically lead to strong spectral correlations between the photons of a created pair and thus to emission into an entangled spectral multimode state \cite{Grice1997,Garay-Palmett2007}. While this state is still pure when emitted, subsequent detection of the pair photons causes the originally coherent superposition of orthogonal broadband spectral modes \cite{Law2000} to collapse into an incoherent mixture. Spectral correlations consequently render the generation of so-called separable photon pair states impossible \cite{U'ren2005}, which strongly affects the photon pair's usefulness for interference experiments and any quantum network application.

By taking one of the photons as an indicator for the presence of the other, photon pair sources can be utilized as sources of heralded single photons \cite{Hong1986} and thus, for example, as basic building blocks for linear optical quantum computing (LOQC) \cite{Knill2001}. However, a high success probability of LOQC schemes can only be achieved if the interacting heralded single photons are indistinguishable and events from different sources can be easily synchronized. Hence the generation of photons in short wave packets---providing accurate timing information---is a crucial component for future quantum circuits.
In the presence of spectral correlations, the common approach for preparing indistinguishable heralded photons is to apply narrowband spectral filtering \cite{Fulconis2007,Chen2006,McMillan2009}. However, this method has two major drawbacks: It significantly reduces the available photon rate and perfect indistinguishability can only be achieved in the limit of infinitely narrow filtering \cite{Branczyk2009}. This, in turn, inevitably leads to photons with narrow spectra and consequently long coherence times, which prevents the generation of ultrashort photon wave packets for quantum applications requiring precise timing.

The only solution to these constraints is a source that directly provides both spatially and spectrally decorrelated photon pairs. Only recently have the first SPDC and SFWM sources been implemented that achieve this task. So far, these sources operate exclusively in either the region from approximately 600 to 900~nm \cite{Mosley2008} or at telecom wavelengths \cite{Kuzucu2008}. As modern optical fiber communication is mostly based in the C telecom band between 1530 and 1565~nm, sources compatible with this wavelength region are clearly needed. At the same time, the best-performing single photon detectors are based on silicon avalanche photodiodes (APDs) and have their peak efficiency between 500 and 850~nm. This results in another considerable advantage of the silicon APD wavelength regime: the availability of efficient photon-number resolving detection schemes \cite{Achilles2003}. Hence, a source that combines the silicon APD region and the telecom band will open up new potentials for many quantum optics applications. While photon pair sources exist that combine both regions \cite{Fasel2004,Rarity2005,McMillan2009}, none of them are intrinsically decorrelated. Consequently, they are not capable of producing pure heralded single photon wave packets.

Here we present a pair source of spectrally decorrelated single photon pulses that bridges the visible and telecommunication wavelength regions. The emission of the signal photon in the visible allows for efficient detection with silicon APDs and the heralding of broadband ultrashort idler photon wave packets in the telecom C band. By increasing the pump power and taking advantage of photon-number resolving detection schemes for the signal, the conditioned generation of higher photon number states at telecom wavelengths is feasible.

\begin{figure}
\includegraphics[width=.48\textwidth]{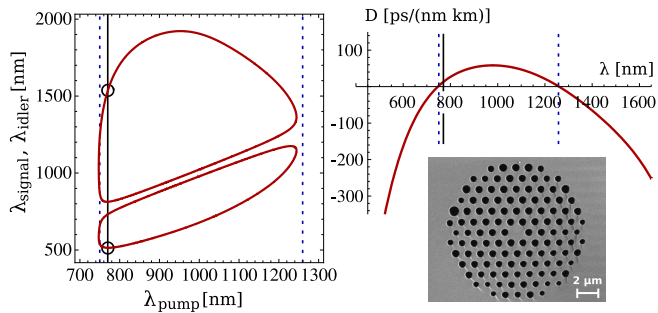}
\caption{\label{fig:pm_plot} Phase-matching plot based on the measured PCF dispersion. At a pump wavelength of 771~nm (black vertical line) the group velocities of idler and pump are matched and the generation of spectrally decorrelated photon pairs with the signal at 514~nm and the idler at 1542~nm is possible. Upper right: Measured dispersion parameter of the fiber with two zero dispersion wavelengths (ZDWs) at 747 and 1260~nm (dashed blue lines). Lower right: Scanning electron micrograph of the PCF with a hole-to-hole distance of 1.2~$\mu$m and a hole diameter of 0.7~$\mu$m.}
\end{figure}

Our source is based on four-wave mixing (FWM) in photonic crystal fiber (PCF), a $\chi^{(3)}$ nonlinear process that allows the spontaneous conversion of two pump photons into a new pair of photons, in general referred to as signal and idler. In addition to energy conservation ($ 2 \omega_p = \omega_s + \omega_i $) the conversion process has to obey the phase-matching condition
\begin{equation}
2 k(\omega_{p}) - 2 \gamma P_p = k(\omega_{s}) + k(\omega_{i}),
\label{eq:pm}
\end{equation}
where the $k(\omega_j) = n(\omega_j) \omega_j/c$ (with $j = p, s, i$) represent the wave vectors of pump, signal, and idler, respectively. The additional phase factor, consisting of the peak pump power $P_p$ and the nonlinear coefficient $\gamma$, is caused by self-phase modulation \cite{Agrawal2007}. Equation~(\ref{eq:pm}) illustrates the importance of dispersion to the FWM process. In order to generate separable photon pairs with specific signal and idler wavelengths, the dispersion of the nonlinear medium has to be designed accordingly. Due to its customizable microstructure, PCF offers the opportunity to engineer a wide variety of dispersion profiles. This has led to the implementation of highly tunable FWM sources both at the single photon level \cite{Rarity2005} and at higher powers \cite{Andersen2004}.

We designed and fabricated a solid-core PCF to be used for the generation of highly nondegenerate spectrally decorrelated photon pairs. The chosen PCF structure yields a group-velocity dispersion profile with one zero dispersion wavelength (ZDW) at 747~nm and a second ZDW at 1260~nm (see Fig.~\ref{fig:pm_plot}). Note that other PCF sources bridging the silicon APD and telecom wavelength ranges have to be pumped in the normal dispersion regime \cite{Rarity2005,McMillan2009}. This fact is imposed by the fibers' dispersion properties and prohibits the generation of spectrally decorrelated pairs due to lack of the required group-velocity balance between pump, signal, and idler. Our source design with two adequately placed ZDWs, however, allows the production of a spectrally decorrelated and thus separable pair state when pumping in the anomalous dispersion regime at 771~nm. The generated idler at 1542~nm then becomes group-velocity matched with the pump, resulting in the emission of signal and idler into a single spectral mode. The corresponding signal is created at 514~nm.

For our PCF, spatial correlations do not have to be taken into account, since pump, signal, and idler all propagate in the fundamental fiber mode. Spectral correlations, however, have to be considered, as they prohibit spectral single mode characteristics and consequently any description as a single pair of single photon wave packets is not valid. This can be understood when looking at the general expression for the photon pair state which can be approximated to first order by
\begin{equation}
|\psi_{s,i}\rangle \propto \int d\omega_{s} d\omega_{i} f(\omega_{s},\omega_{i}) a^{\dagger}(\omega_s) a^{\dagger}(\omega_i) |0\rangle.
\end{equation}
Only when the relative group velocities of pump, signal, and idler are adapted appropriately \cite{Garay-Palmett2007} can the spectral distribution function $f(\omega_{s},\omega_{i})$ be decomposed into a product $ f(\omega_s,\omega_i) = f_1(\omega_s) f_2(\omega_i) $, which, in turn, constitutes a separable state $ |\psi_{s,i}\rangle = |\psi_{s}\rangle \otimes |\psi_{i}\rangle $.

\begin{figure}
\includegraphics[width=.4\textwidth]{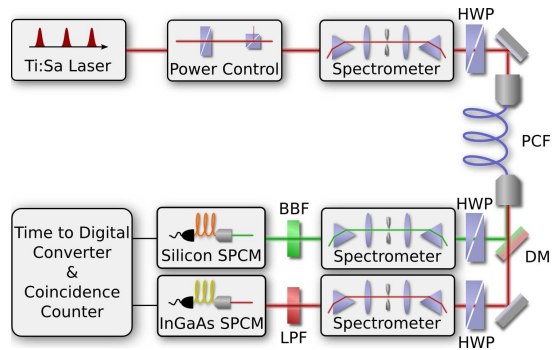}
\caption{\label{fig:setup} Schematic of the setup for measuring the spectral coincidence distribution of our source (see text for details).}
\end{figure}

After measuring the dispersion and structure of our PCF, we first set up an experiment to analyze the spectral properties of the generated photon pairs (see Fig.~\ref{fig:setup}). Our Ti:sapphire femtosecond pump pulses were restricted to a bandwidth of 3~nm by a 4f prism spectrometer and then launched into 65~cm of fiber. Earlier analyses had shown that our PCF was weakly birefringent ($\Delta n \approx 10^{-4}$), the two polarization axes exhibiting slightly different phase-matching conditions. A half-wave plate (HWP) at the input coupling end of the PCF ensured pumping on one of the major axes. At the output, signal and idler were separated by a dichroic mirror (DM) and spectrally resolved with prism spectrometers. Half-wave plates in front of each spectrometer ensured transmission with minimal Brewster losses. A broadband filter in the signal arm (BBF, Schott BG39) and longpass filters in the idler arm (LPF, Semrock LP02-808RS-25 and LP02-1064RS-25) suppressed the residual pump and stray light before detection. The signal was then coupled into a multimode fiber and detected by a silicon single photon counting module (SPCM) (Perkin-Elmer AQR-14-FC), the idler was coupled into a single-mode fiber, and an InGaAs SPCM (idQuantique id201) was used for detection. Both detector outputs were connected to a time-to-digital converter. By independently scanning the two motorized spectrometer stages we were thus able to measure the spectral coincidence distribution of our source.

\begin{figure}
\includegraphics[width=.42\textwidth]{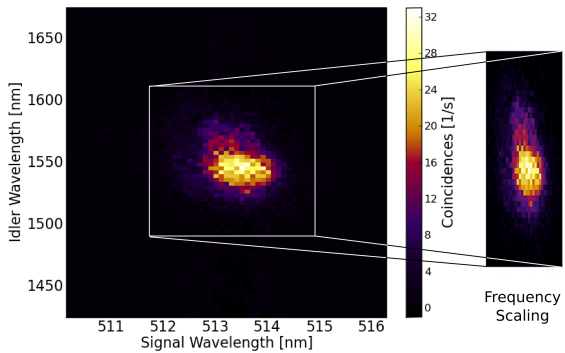}
\caption{\label{fig:jsi} Joint spectral coincidence measurement for a pump wavelength of 771~nm with 250~$\mu$W average pump power at a repetition frequency of 1.5~MHz. As signal and idler axes span different frequency ranges, a rescaled plot is included in which both axes are scaled identically.}
\end{figure}

The obtained joint spectral coincidence distribution at a pump wavelength of 771~nm is shown in Fig.~\ref{fig:jsi}. The central signal wavelength was found to be 513.5~nm and the central idler wavelength 1547~nm, values which agree well with the expected wavelengths displayed in Fig.~\ref{fig:pm_plot}. Correcting the measured full width at half maximum (FWHM) of 1.4 and 42.5~nm for the limited resolution of our spectrometers (0.7~nm in the signal and 24~nm in the idler arm), we calculated the signal marginal FWHM to be 1.2~nm and the idler marginal FWHM to be 35~nm.

For quantifying the correlations present in the spectrum of a given state, one possibility is to apply a Schmidt decomposition. The spectral distribution function can be decomposed into the sum
$f(\omega_{s},\omega_{i}) = \sum_{n=0}^{N-1} \sqrt{\lambda_n} \xi_n^{(1)}(\omega_{s}) \xi_n^{(2)}(\omega_{i}),$
with $\{\xi_n^{(1)}(\omega_{s})\}$ and $\{\xi_n^{(2)}(\omega_{i})\}$ each forming a set of orthonormal functions with broad spectra, the Schmidt modes. The Schmidt coefficients $\lambda_n$ are real valued and fulfill $\sum_n \lambda_n = 1$. The Schmidt decomposition is a valuable tool for deriving spectral entanglement from joint spectral intensity measurements \cite{Law2000}. The parameter $K = 1/\sum_n \lambda_n^2$ is directly related to the purity $P$ of the heralded single photons derived from the pair state via $P = 1/K$. Applying a singular value decomposition, the numerical analog of the Schmidt decomposition \cite{Mauerer2008}, to our recorded joint spectral coincidence measurement yields $K = 1.22$. This corresponds to a purity of our heralded single photons of up to 82\%. The result shows that our source emits spectrally decorrelated photon pairs of high quality without any need for narrowband filtering. It also implies that the complete idler spectrum of 35~nm FWHM can be used in future experiments and applications and that the generation of ultrafast photon wave packets becomes feasible. The spectral width ideally corresponds to the generation of photon wave packets at 1547~nm with a transform limited duration of 100~fs.

In order to characterize the source brightness, we set up an experiment with a dichroic mirror to separate the pair photons, but this time coupling both signal and idler directly into multi- and single-mode fibers, respectively. As we had already proven that narrowband filtering is unnecessary, only broadband and edgepass filters were used to suppress pump and stray light. At an average pump power of 100~$\mu$W and a repetition rate of 1~MHz we measured a raw coincidence rate of $1.3\times10^3$ counts/s with signal and idler count rates of $16.5\times10^3$ and $6.0\times10^3$, respectively. The analysis of coincidences between subsequent pulses, a typical way of estimating dark count and higher photon number contributions, yielded 100 counts/s. Subtracting this number from the raw coincidence rate results in $1.2\times10^3$ coincidences per second and signal and idler detection efficiencies of 20\% and 7\%, respectively. Note that these detection efficiencies include all coupling and filtering losses as well as the detector efficiences (45\% for the Si SPCM, 25\% for the InGaAs SPCM). The efficiency $\eta_h$ of heralding our idler wave packets can be calculated to be 28\% via $\eta_h = R_c/(R_s \eta_{\text{InGaAs}})$, where $R_c$, $R_s$, and $\eta_{\text{InGaAs}}$ denote the coincidence rate, signal rate, and InGaAs SPCM detection efficiency, respectively.

\begin{figure}
\includegraphics[width=.37\textwidth]{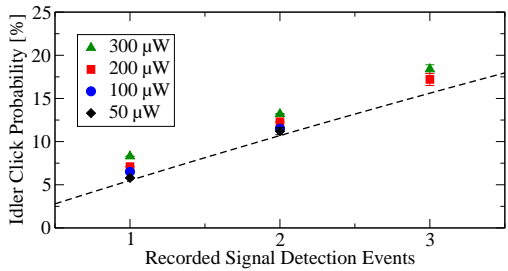}
\caption{\label{fig:probs} Idler click probabilities for several pump powers as a function of signal detection events $n$. A theoretical curve given by $ P_{\text{click}}(n) = 1 - (1 - \eta_i)^n $ based on the measured idler detection efficiency of $\eta_i = 5.5\%$ is included. The offset between measured data and theory, especially for high pump powers, is due to the contribution of even higher photon number components. As no significant three-click rate was present at the two lowest pump powers, these data are not included.}
\end{figure}

Another intriguing property of our source is the possibility to generate higher photon numbers at relatively moderate pump powers. To prove this, we performed measurements with a time-multiplexing detector (TMD) \cite{Achilles2003} in the signal arm and a single InGaAs SPCM in the idler arm. The TMD allows the detection of up to eight photons in a pulse using only two silicon SPCMs by splitting the original input pulse into several time-delayed subpulses. In the analysis of our measured data we were thus able to condition on multiple click events in the signal arm and to observe the corresponding detection event probability, that is, click probability of the SPCM, in the idler arm. If the overall detection probability of one photon in the idler arm is defined as $\eta_i$, the probability of registering a detector click when $n$ photons arrive is given by $P_{\text{click}} = 1-(1-\eta_i)^n$. Thus the click probability $P_{\text{click}}$ increases with higher photon numbers and even a single detector can provide limited information about photon number. Figure~\ref{fig:probs} shows our recorded idler click probabilities for different pump powers when we conditioned on one, two or three detection events with the TMD in the signal arm. The increase in idler click probability when conditioned on multiple clicks in the signal arm is a clear indication that the corresponding higher photon numbers are also present in the idler arm. The recorded idler click probabilities increase according to theory for small pump powers, but show a noticeable offset at higher pump powers. The offset is caused by imperfect preparation of the photon number states due to the contribution of higher photon number components. Our complete output state can be described by $|\psi\rangle = \sum_{n=0}^{\infty} \sqrt{P_n} |n,n\rangle$, where $P_n$ represents the probability of finding $n$ photons in both the signal and idler arms. $P_n$ depends on the mean photon number of the state, which, in turn, is a function of the pump power. Consequently, the probability of generating higher photon numbers increases with higher pump powers and leads to an increased click probability. The fact that $|\psi\rangle$ also includes photon number components greater than the number of signal clicks upon which the measurement was conditioned causes an additional increase of the idler click probability and thus the observed offset in Fig.~\ref{fig:probs}.

Due to the additional coupling to the time-multiplexing fiber network the overall signal detection efficiency was reduced to 9.4\% in this experiment. Nevertheless we were able to detect up to 40 two-click signal events per second at 100~$\mu$W average pump power. At a pump power of 300~$\mu$W we recorded approximately 2200 two-click, 40 three-click, and 0.5 four-click events per second in the signal arm. This result shows that our source offers the possibility to provide heralded Fock states with higher photon numbers at telecom wavelengths.

To summarize, we have presented a source of spectrally decorrelated photon pairs that bridge the visible and telecom wavelength regions. An analysis of the joint spectral coincidence measurement results in a photon purity of up to 82\%. This means our source is capable of heralding broadband ultrafast single photon wave packets at telecom wavelengths. We have further shown that by increasing the pump power, the efficient generation and thus conditional preparation of higher photon numbers in this wavelength regime are feasible. We believe that this source will be beneficial for many quantum-enhanced technologies, for example, quantum imaging \cite{Giovannetti2004}, quantum computing \cite{Knill2001}, and quantum cryptography \cite{Mauerer2007a}.

This work was supported by the EC under the grant agreement CORNER (FP7-ICT-213681).

\end{document}